\documentclass[12pt]{iopart}
\usepackage{graphicx}

\begin{document}

\title[C$_{60}$ catalysing H-release in LiBH$_4$]
{Theoretical study of C$_{60}$ as catalyst for dehydrogenation in LiBH$_4$}

\author{Ralph H. Scheicher$^1$, Sa Li$^2$, C. Moyses
Araujo$^1$\footnote{Present address: Department of Chemistry, Yale University, P.O.\ Box 208107, New Haven, Connecticut 06520-8107, USA},
Andreas Blomqvist$^1$\footnote{Present address: Sandvik Tooling R\&D, SE-126 80 Stockholm, Sweden}, Rajeev Ahuja$^{1,3}$ and Puru Jena$^2$}

\address{$^1$ Condensed Matter Theory Group,
Department of Physics and Astronomy, Box 516, Uppsala
University, SE-751 20 Uppsala, Sweden}

\address{$^2$ Physics Department,
Virginia Commonwealth University, Richmond, Virginia 23284, USA}

\address{$^3$ Applied Materials Physics,
Department of Materials and Engineering, Royal Institute of
Technology (KTH), SE-100 44 Stockholm, Sweden}

\ead{ralph.scheicher@physics.uu.se}

\begin{abstract}
Complex light metal hydrides possess many properties which make them
attractive as a storage medium for hydrogen, but typically,
catalysts are required to lower the hydrogen desorption temperature
and to facilitate hydrogen uptake in the form of a reversible
reaction. The overwhelming focus in the search for catalysing agents
has been on compounds containing titanium, but the precise mechanism
of their actions remains somewhat obscure. A recent experiment has
now shown that fullerenes (C$_{60}$) can also act as catalyst for
both hydrogen uptake and release in lithium borohydride (LiBH$_4$).
In an effort to understand the involved mechanism, we have employed
density functional theory to carry out a detailed study of the
interaction between this complex metal hydride and the carbon
nanomaterial. Considering a stepwise reduction of the hydrogen
content in LiBH$_4$, we find that the presence of C$_{60}$ can lead
to a substantial reduction of the involved H-removal energies. This
effect is explained as a consequence of the interaction between the
BH$_x^-$ complex and the C$_{60}$ entity.
\end{abstract}


%
%
%
%

\submitto{\NT}

\maketitle

\section{Introduction}

Hydrogen possesses many features which make it a highly attractive
option as a future energy carrier \cite{Schlapbach:2001,Schlapbach:2009,LiuChang:2010a,Jena:2011}. While such
a transition is certainly not going to be easy, it is nonetheless
inevitable to explore feasible alternatives to our fossil fuel
resources which are finite and might even become exhausted within
this century, and which furthermore involve carbon dioxide emissions
linked to a harmful global climate change. Before one could even
begin to realize such an envisioned hydrogen economy, a number of formidable
technological problems need to be solved first. One of these issues
is hydrogen storage \cite{Schlapbach:2001,Schlapbach:2009,LiuChang:2010a,Jena:2011}. Pressurized gas tanks and cryostatic storage of
hydrogen as a liquid do not seem ideal for mobile (vehicular)
applications. Rather, storage of hydrogen inside solid materials
appears to be the most feasible solution.

In this context, complex light metal hydrides are very promising as
a suitable hydrogen storage medium. On the plus side, the
gravimetric hydrogen content in these materials can be quite large,
but a significant drawback is the generally high temperature
required for hydrogen desorption and the fact that the reactions are
often not reversible, i.e., hydrogen can be released, but the direct
rehydrogenation is difficult or even impossible. To address
this issue, catalysts are commonly added to the metal hydride
systems, and particularly titanium has been shown to lower hydrogen
desorption temperatures and enable reversibility
\cite{Bogdanovic:1997}, albeit the precise mechanism remains a topic
of debate (for a discussion, see Refs.\ \cite{Araujo:2005a,Araujo:2005b}).

Studies of transition metal-catalysed hydrogen storage properties of sodium alanate (NaAlH$_4$) went beyond Ti, and also considered Sc, Zr, and other early transition metals \cite{Zidan:1999,Sandrock:2002,Anton:2002,Bogdanovic:2006,Sakintuna:2007,Blomqvist:2007}. Still, a fundamental understanding of how these catalysts, and in particular Ti, function has not been
achieved yet \cite{Balde:2007,Haiduc:2005,Majzoub:2003,Leon:2006}.
It is widely believed
that a better insight into how a catalyst works
could facilitate the rational design of novel catalysts for
alanates, borohydrides, and amides.

Carbon nanomaterials have recently been added to the list of
materials that could potentially act as ``true catalysts'' in the
sense that they retain their structure and are not consumed in the
reaction, and furthermore would be conceptually much simpler to
understand in their action \cite{Cento:2007,Dehouche:2005,Zaluska:2000,Pukazhselvan:2005,Wang:2006,Berseth:2009}.
Besides this function as catalysts \cite{LiuChang:2010b,Amirkhiz:2010,Hudson:2010,WuChengzhang:2010,Teprovich:2011,Ismail:2011,Raghubanshi:2011,Hudson:2011}, carbon nanomaterials have also been considered for their role in nanoconfinement \cite{WuChengzhang:2010,Verkuijlen:2010,GaoJinbao:2010,Adelhelm:2010,deJonghPetra:2010,Ngene:2010,Li:2011,Jensen:2011} to aid in the hydrogen desorption process.

Previously, we presented the results of
our study combining experiment and theory to demonstrate and understand the
catalytic effects of various carbon nanomaterials on the hydrogen
release and uptake in NaAlH$_4$ \cite{Berseth:2009}. We found that the electron affinity of the studied carbon nanomaterials is mainly determined by the surface curvature and directly affects the hydrogen sorption mechanism of NaAlH$_4$. One mechanism identified by us considers the higher electron affinity causing Na to donate more of its electronic charge to the carbon substrate, thus destabilizing the AlH$_4$ complex, which ultimately leads to a reduction in the H-removal energy. More recently, the
experimental studies were expanded to also cover hydrogen sorption in LiBH$_4$, for which
it was found again that carbon nanomaterials, and in particular
fullerenes (C$_{60}$), are excellent in their function as a catalyst
\cite{Wellons:2009}.

Hydrogen desorption from LiBH$_4$ takes place in two steps:
\begin{equation}
2 \textrm{LiBH}_4  \rightarrow  2 \textrm{LiH} + 2 \textrm{B} + 3 \textrm{H}_2
\label{reaction1}
\end{equation}
\begin{equation}
2 \textrm{LiH}  \rightarrow 2 \textrm{Li} + \textrm{H}_2
\label{reaction2}
\end{equation}
Step \ref{reaction1} occurs above 400$^\circ$C and releases 13.8 wt\% hydrogen, while
step \ref{reaction2} releases 18.4 wt\% hydrogen and takes place at about 950$^\circ$C which is too high a temperature to be considered practical (see Ref.\ \cite{Wellons:2009} and references therein).

Here, we present the results from our first-principles analysis of the catalysing mechanism that enables C$_{60}$ to enhance the hydrogen desorption from LiBH$_4$. In our previous work on NaAlH$_4$, we had only considered the removal of a single hydrogen atom \cite{Berseth:2009}, but in the present study, we went further and explored the stepwise removal of up to three hydrogen atoms from LiBH$_4$. This approach has led us to identify some new features in the
C$_{60}$-catalysed desorption reactions of LiBH$_4$ as compared to NaAlH$_4$.

\section{Computational methods}
Total energy calculations were carried out within the framework
of the generalized gradient approximation (GGA) \cite{Perdew:1992} to
density functional theory \cite{Hohenberg:1964,Kohn:1965} by using
the projector augmented wave (PAW) method \cite{Blochl:1994} as
implemented in the Vienna Ab initio Simulation Package (VASP)
\cite{Kresse:1996}. PAW potentials with the valence states $1s$ for H, $2s$ for Li, $2s2p$ for B, and
$2s2p$ for C were
employed. The calculations under periodic-boundary conditions utilized a cubic supercell box with a
side length of 20~\AA, which prevented any unphysical interaction between repeated images and allowed us to consider only the
$\Gamma$-point for Brillouin zone sampling. Atomic positions were
relaxed with respect to minimum forces using a conjugate-gradient
algorithm. To ensure that we identify indeed the ground-state structure, we initiated the relaxation process from several different starting configurations which saw the LiBH$_x$ unit (for $x$ = 4, 3, 2, and 1) located at various potential trapping sites on the fullerene surface (such as on top of a C atom, on a C--C bridge, or a hollow site) and in various orientations.

\section{Results and discussion}

\noindent
\textbf{BH$_{4-x}$$^{0/-}$}

\begin{figure}
\begin{center}
    \includegraphics[width=10.0cm]{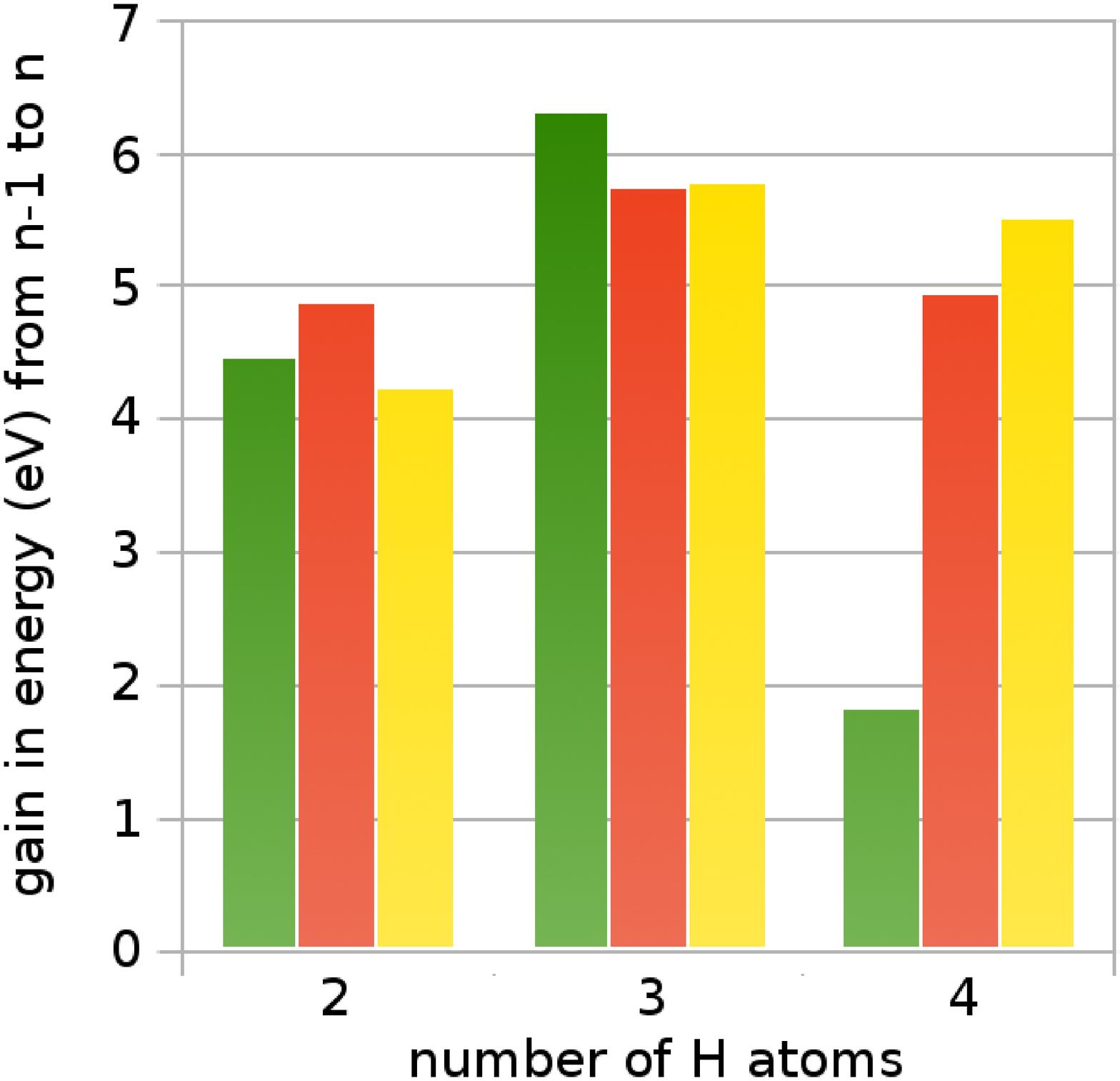}
    \caption{This plot shows the energy gain when adding a hydrogen atom to the respective clusters of
    BH$_{n-1}$ (left-side green bars), BH$_{n-1}^-$ (central red bars), and LiBH$_{n-1}$ (right-side yellow bars).} \label{fig:energy_gain}
\end{center}
\end{figure}

Before introducing C$_{60}$, we calculated the energetics for isolated BH$_{4-x}$ units
($x$ = 0, 1, 2, and 3), both in the charge-neutral state (BH$_{4-x}$$^0$) and with
one extra electron added (BH$_{4-x}$$^-$), and for the corresponding
isolated LiBH$_x$ units. From this data, we calculated the gain in energy when a hydrogen atom is added and compared the results within each of these three systems, i.e., charged and neutral BH$_{4-x}$ and LiBH$_x$. The findings from this analysis are summarized in figure
\ref{fig:energy_gain}. Quite analogous to our earlier published
analysis for AlH$_{4-x}$$^{0/-}$ \cite{Berseth:2009}, it is seen that
the largest stabilization due to the extra electron occurs for the complex containing four hydrogen atoms, i.e., BH$_4$.

We now turn to the main part of our investigation, namely the stepwise dehydrogenation in LiBH$_4$ when supported on C$_{60}$. In the following, we present and discuss the results from our
density functional calculations for each of the LiBH$_{4-x}$+C$_{60}$
entities where $x$ takes on the values 0, 1, 2, and 3.

\noindent
\textbf{LiBH$_4$}

\begin{figure}
\begin{center}
    \includegraphics[width=10.0cm]{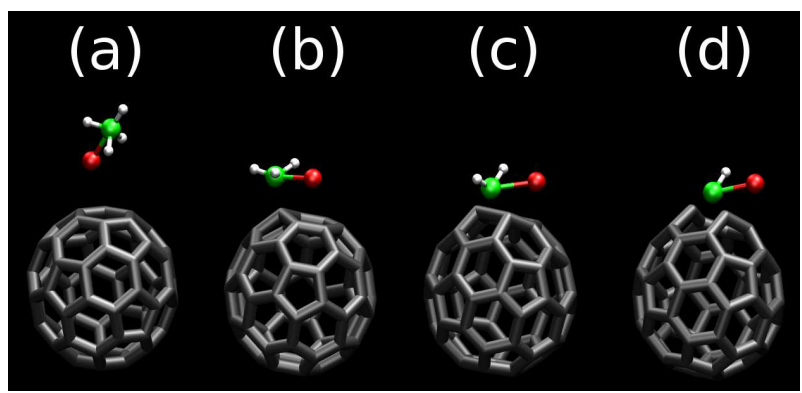}
    \caption{Step-wise dehydrogenation of LiBH$_4$ catalysed by C$_{60}$. Shown here are the atomic configurations as obtained from \emph{ab initio} geometry
optimizations for fullerene plus (a) LiBH$_4$, (b) LiBH$_3$, (c) LiBH$_2$, and (d)
LiBH. Li is colored in red, B in green, H in white, and C in gray.}
\label{fig:structures}
\end{center}
\end{figure}

The most stable configuration has Li of the
LiBH$_4$ unit centered above a C--C bridge (figure
\ref{fig:structures}a). Other configurations which had LiBH$_4$ located at different sites on C$_{60}$ were tested, but were found to yield higher energies. For example, a 90$^\circ$-rotation of the
LiBH$_4$ unit (so that both Li and B would have roughly comparable
distances to the fullerene surface) results in a drastic increase in energy
by nearly 1 eV. Also, placing the LiBH$_4$ unit directly above the
center of the hexagon yields a state about 0.25 eV higher in energy.
The binding energy of LiBH$_4$ to C$_{60}$ is calculated as 0.50 eV, implying that the interaction is rather weak. 
A Bader analysis of the charge density shows that lithium has
completely lost its electron. But this charge does not appear to
have been transferred to the fullerene; rather, LiBH$_4$ as a whole
is charge-neutral, so here the electron seems to have been used to
stabilize the BH$_4$$^{-}$ unit, in line with what one would expect based on the results from our analysis of the
energetics of the charged and neutral BH$_x$ units above.
In addition, partial density of states (PDOS) calculations show that the B--H bonds in the BH$_4$$^{-}$ unit are of covalent nature, while the interaction between Li$^+$ and BH$_4$$^{-}$ is of ionic nature.

\noindent
\textbf{LiBH$_3$}

\begin{figure}
\begin{center}
    \includegraphics[width=10.0cm]{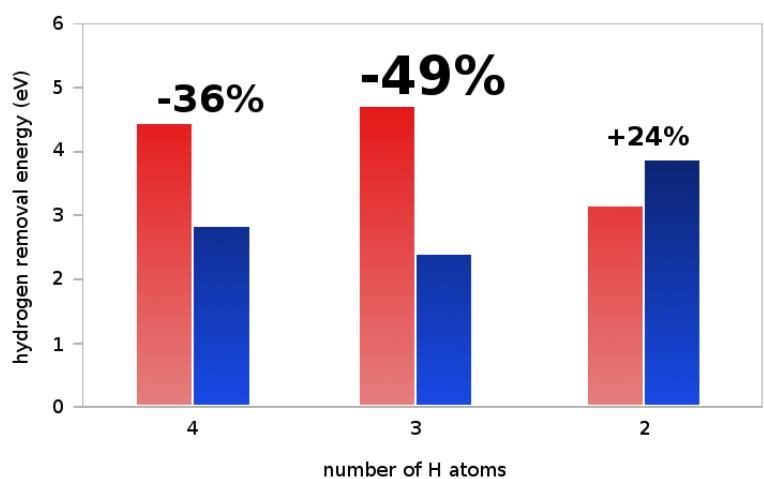}
    \caption{The hydrogen removal energies are plotted as a function of hydrogen content
    in LiBH$_x$ (from left to right for $x$ = 4, 3, and 2 H atoms), in the absence of fullerene (left-side red bars) and when fullerene is present (right-side blue bars). The relative change in hydrogen removal energy due to presence of the C$_{60}$ catalyst is printed above the bars.} \label{fig:removal_energetics}
\end{center}
\end{figure}

Upon removal of the first hydrogen atom, we arrive at LiBH$_3$ supported on C$_{60}$.
Remarkably here, a configuration of the LiBH$_3$
unit similar to the upright one for LiBH$_4$ is not the most stable
configuration. Instead, a ``sideways'' orientation (figure
\ref{fig:structures}b) yields the ground state (0.86 eV lower in
energy than the ``upright'' orientation). This enables the boron of
the BH$_3$ unit to interact with one carbon atom of the fullerene,
thus allowing it to preserve an approximate tetrahedral structure in
the form of BH$_3$C. One can even notice that the carbon atom has
been slightly ``pulled out'' by the boron, distorting the hexagon
somewhat. Indeed, PDOS calculations show that a bond of covalent nature is formed between B and this particular C atom. 
In a sense, this interaction of BH$_3$ with C$_{60}$
compensates for the loss of a hydrogen and substantially lowers the
energy, as can also be seen from the greatly enhanced binding energy
of LiBH$_3$ on C$_{60}$ (2.09 eV), and this plays a crucial role for the low
H-removal energy of 2.82 eV in the first step (figure
\ref{fig:removal_energetics}). However, it should be noted that the actual energy cost for removing the hydrogen atom is higher than that, due to a reaction barrier which was calculated by us with the nudged elastic band method \cite{Jonsson:2000} to be 4.3 eV high.
As for the charge state, it is seen
that lithium retains about 10\% of its electron, but the LiBH$_3$
unit as a whole is missing $-$0.7 e, so here a substantial charge
transfer to the C$_{60}$ has indeed taken place. In fact, the carbon
atom closest to boron carries an excess $-$0.37 e. The rest of the
charge is distributed over the other 59 carbon atoms of the
fullerene. Thus, a complex cooperative process is at work here, in which the extra electron that stabilized the BH$_4$ unit is no longer needed in BH$_3$ following the H atom removal, and can now instead facilitate the formation of a bond between B and C.

\noindent
\textbf{LiBH$_2$}

Removal of the second hydrogen atom leads to LiBH$_2$ which just like LiBH$_3$ prefers a ``sideways''
orientation that is found to be more stable than the ``upright''
orientation (figure \ref{fig:structures}c). The difference in energy
between these two possible configurations amounts in this case even
to 3.5 eV, apparently owing to the now possible stronger interaction
of boron with two carbon atoms of the fullerene. The binding energy of LiBH$_2$ to C$_{60}$ is 4.40 eV, about twice as much as in the case for LiBH$_3$, reflecting that B interacts with two C atoms now. This is also seen from the PDOS, showing the formation of two covalent bonds between B and the two C atoms. 
The tetrahedral structure is again approximately preserved in the BH$_2$C$_2$
configuration that is formed. Thus the two carbon atoms act again as
substitutes for the removed hydrogen atoms and are seen to be
``pulled up'', away from the C$_{60}$, towards the boron, without actually breaking the C$_{60}$ framework.
The lowering of the total energy through this form of interaction
between BH$_2$ and the fullerene is once again responsible for the
even further lowered H-removal energy of 2.38 eV in the second step
(figure \ref{fig:removal_energetics}). The charge state situation is
somewhat similar to the previous case of LiBH$_3$, in the sense that
lithium is charged positively with +0.88 e, but the LiBH$_2$ now
carries an even higher positive charge of +1.4 e, twice as much as
in the previous case. Of course, here, boron is interacting with two
carbon atoms and indeed it is seen that they both receive about
$-$0.5 e in excess charge.

\noindent
\textbf{LiBH}

After the third hydrogen atom has been removed, it is
no longer possible for boron to easily preserve the preferred
tetrahedral arrangement of binding partners as seen in the previous
three cases. Instead, the BH unit continues to interact with two
carbon atoms and only to a weaker extent with lithium (figure
\ref{fig:structures}d). As a consequence, the H-removal energy in
the third step is found to be not lowered, but quite oppositely,
increased with respect to the situation of an isolated LiBH
unit where no C$_{60}$ is present (figure
\ref{fig:removal_energetics}). 
The binding energy of LiBH to C$_{60}$ amounts to 3.66 eV, similar to the value for LiBH$_2$, due to both having B effectively interact with two C atoms, but reduced perhaps because of the strained non-tetrahedral configuration in which B finds itself. The PDOS again shows the formation of two covalent bonds between B and the two C atoms. 
We also tested the possibility of H
attaching to Li rather than B, but the relaxation results indicate
that H does indeed prefer to bind to B in this cluster.

\section{Concluding remarks}

Carbon nanomaterials certainly show great promise as catalysts for hydrogen sorption in complex light metal hydrides. In the present theoretical study, we have concentrated on the fullerene C$_{60}$ acting on the borohydride LiBH$_4$, motivated by definitive experimental evidence for the catalysing effects in this particular combination \cite{Wellons:2009}. But as we had shown earlier \cite{Berseth:2009}, a similar outcome was seen for the alanate NaAlH$_4$, and the carbonaceous catalysing agent was found to be not limited to C$_{60}$, but could for example as well be high-curvature single-walled carbon nanotubes. Electron affinity certainly plays an important role, but not necessarily always the same role. For NaAlH$_4$ on various carbon substrates, we had found a charge transfer which presumably destabilized the AlH$_4$ complex \cite{Berseth:2009}. Contrary to that, we did not find any theoretical evidence in our calculations here for charge transfer occurring from LiBH$_4$ to C$_{60}$. Instead, the mechanism, which led to a sizeable reduction in the energy required to remove the first hydrogen atom in LiBH$_4$ relies on an increased stabilization of the product state, namely by formation of a ``substitutional'' bond between the B atom of LiBH$_3$ and a C atom of C$_{60}$. That bond is however facilitated by a charge transfer reaction; hence electron affinity comes into the equation again. An analogous mechanism, now involving two C atoms, is responsible for the even further reduced hydrogen removal energy when going from LiBH$_3$ to LiBH$_2$. Those are the main findings of this theoretical investigation.

Obviously, there are ways in which the present study could be improved. For example, our cluster approach only considered a single LiBH$_4$ unit interacting with C$_{60}$. Although often, such minimal clusters mirror many of the important qualitative properties of the parent solid that they stand for \cite{Jena:2010}, it might still be worthwhile to expand the cluster to include more than one LiBH$_4$ unit. Of course, this will significantly raise the complexity of the investigation, but could be the only way to identify overlooked mechanisms in which a cooperative interaction between several LiBH$_4$ units on the surface of a C$_{60}$ leads to a reduction in hydrogen removal energy. Certainly, there are many more questions to be answered before we have a complete understanding of the catalysing effects of carbon nanomaterials on complex light metal hydrides, and we hope that the present work has not merely contributed some new insights to this field, but, perhaps more importantly, will also stimulate fresh theoretical and experimental investigations.

\ack RHS, CMA, AB, and RA thank the Swedish Research Council and Futura Foundation for financial support. SL and PJ thank the U.S.\ Department of Energy and the Office of Basic Energy Science for funding. Computational facilities for this project were provided by the Swedish National Infrastructure for Computing (SNIC) and by the Uppsala Multidisciplinary Center for Advanced Computational Science (UPPMAX).

\newpage

\end{document}